\documentclass[journal,10]{IEEEtran}

\usepackage{graphicx}
\usepackage{url}
\usepackage{enumitem}
\usepackage{booktabs}
\usepackage{color}
\usepackage{subfig}
\usepackage{amsmath}

	\title{TRUAV: Distributed Multi‑Agent Reinforcement Learning for Trajectory Planning and Routing Enhancement in UAV‑Aided IoT‑Enabled VANETs}

\author{Muhammad Umar Farooq Qaisar, Lin Zhang, Zhen Chen, Wajdy Othman, Shehzad Ashraf Chaudhry, and Chang Liu
	\thanks{M. U. F. Qaisar (muhammad@buaa.edu.cn) is with the Hangzhou International Innovation Institute of Beihang University, Hangzhou 311115, China}
	\thanks{L. Zhang (zhanglin@buaa.edu.cn) is with Hangzhou International Innovation Institute of Beihang University, Hangzhou 311115, China, the School of Automation Science and Electrical Engineering at Beihang University, Beijing 100191, China, and the State Key Laboratory of Intelligent Manufacturing Systems Technology, Beijing 100854, China}
	\thanks{Z. Chen. (108201001@seu.edu.cn) is currently a Lecturer with the School of Intelligent Science and Engineering and the School of Transportation, Southeast University, Nanjing 211189, China.}
	
	\thanks{W. Othman. (wajdy@nankai.edu.cn) is with the Digital Health Research Center, Haihe Lab of ITAI, jointly with the School of Cyber Science, Nankai University, and China-SCO Digital Intelligence Applications (DIA) Joint Laboratory, Tianjin, China.}
	
	\thanks {S. A. Chaudhry (shehzad.ashraf@adu.ac.ae) is with the Department of Computer Science and Information Technology, College of Engineering, Abu Dhabi University, Abu Dhabi, UAE; and with the Department of Software Engineering, Faculty of Engineering and Architecture, Nisantasi University, Istanbul, Turkey}

	\thanks{C. Liu (liuchang@gdut.edu.cn) is with the School of Information Engineering, Guangdong University of Technology, Guangzhou 510006, China}
	\thanks{Corresponding authors: Muhammad Umar Farooq Qaisar, Email: muhammad@buaa.edu.cn}}

\begin{document}
	\maketitle
	
	\begin{abstract}
		Unmanned aerial vehicles (UAVs) have emerged as a key enabler of next‑generation Internet of Things (IoT) ecosystems, offering flexible aerial relaying to extend connectivity across dynamic vehicular ad hoc networks (VANETs) in smart city environments. However, conventional centralized approaches for UAV trajectory planning require continuous global network state aggregation, making them impractical under bandwidth and energy constraints typical of dense urban deployments. In this article, we present TRUAV, a distributed multi‑agent reinforcement learning framework based on independent tabular Q‑learning for joint UAV trajectory planning and routing enhancement in UAV‑aided VANETs. Each UAV is equipped with a local Q‑learning agent that operates purely on locally observable information, including vehicle density, packet queue states, and neighbor UAV positions, thereby eliminating the need for global state exchange. A potential‑game‑inspired reward design encourages spatial diversity and routing‑aware UAV positioning among interacting agents while accounting for energy consumption. Numerical simulations over a large urban area with 200 mobile vehicles show that the proposed TRUAV framework achieves network coverage and packet delivery ratios comparable to centralized deep reinforcement learning methods, while also improving relay delay and energy efficiency. Finally, we discuss emerging challenges and future research directions for distributed multi‑agent UAV‑assisted IoT systems.
	\end{abstract}

	
\section{Introduction}
\label{sec:intro}

The rapid advancement of intelligent transportation systems has elevated vehicular ad hoc networks (VANETs) to a critical infrastructure component for modern urban mobility. VANETs enable real-time information exchange among vehicles and roadside infrastructure, supporting applications ranging from collision avoidance and cooperative adaptive cruise control to traffic congestion management and passenger infotainment~\cite{1.1}. The underlying communication technologies, including IEEE 802.11p dedicated short-range communications and cellular vehicle-to-everything (V2X) standards, provide the protocol foundations for these services. Yet the dynamic nature of vehicular environments, characterized by high-speed node mobility, frequent topology changes, and intermittent wireless connectivity due to buildings, terrain, and atmospheric conditions, continues to pose fundamental challenges for maintaining robust end-to-end communication in urban settings~\cite{1.2}.


In emerging smart cities, vehicular networks are becoming an integral part of the IoT ecosystem. Vehicles, roadside units (RSUs), and aerial relays can be regarded as mobile IoT nodes that continuously sense, process, and share context information with edge or cloud platforms. UAVs serve as flexible IoT gateways that extend connectivity beyond inflexible physical infrastructure, provide broader situational awareness, and supply additional capacity on demand when traffic load surges in confined areas.

UAVs have emerged as aerial communication platforms and show substantial potential for enhancing VANET connectivity. UAVs typically fly at altitudes between 80 and 150 m, which gives them a distinctive geometric advantage over land-based infrastructure. As a result, they provide significantly better line-of-sight (LoS) propagation conditions for air-to-ground wireless links. This aerial perspective allows the UAVs to maintain reliable communication with ground vehicles that may be blocked by urban structures or located far from fixed RSUs~\cite{1.3}. UAVs acting as flying relays can bridge connectivity gaps and extend network coverage to underserved areas. Moreover, they can be deployed when fixed infrastructure is too expensive or practically infeasible. The collaborative operation of multiple UAVs forms a swarm that aggregates these benefits and creates an aerial IoT layer that is spatially and temporally flexible.


The research community has devoted immense effort into optimizing the UAV’s deployment and movement strategies for vehicular network support. Initial investigations considered static placement problems, in which optimal relay locations are determined to maximize coverage over the ground while keeping the number of UAVs, altitude, and transmit power in check~\cite{2.1}. Later research examined whether designing dynamic trajectories is better than using fixed trajectories or simple reactive movement strategies, considering that the state of the road network is described by a time-varying distribution of vehicles~\cite{1.5}. In recent times, researchers have begun using deep reinforcement learning (DRL) for UAV trajectory optimization in order to automatically learn flight control policies that dynamically adapt to complex and stochastic environment dynamics~\cite{1.6, 1.7}.

In spite of significant recent progress, an important gap remains in the joint treatment of UAV trajectory control and  vehicular IoT routing. Many existing approaches decouple trajectory planning from routing optimization: UAVs are positioned to maximize aerial coverage (or signal quality), without explicit consideration of packet source, packet destination and which aerial relays can most effectively forward packets. This disconnection results in a disparity between UAV locations and the communication requirements, thus undermines the quality of service for delay-sensitive and data-intensive vehicular applications.

Chen et al.~\cite{1.8} address this issue using a centralized multi-agent soft actor-critic framework that jointly optimizes the UAV trajectories and VANET routing with the help of a global critic network. Their solution achieves throughput gains of approximately 20 to 35 percent over static UAV placement. Nonetheless, the centralized critic demands full state knowledge of all agents at each training step, incurring communication overhead that scales quadratically with the swarm size. With six UAVs associated with 50-dimensional state vector at 10 Hz, the controller processes 15000 values a second. This creates a communication bottleneck and single point of failure that becomes prohibitive as deployments scale. UAVs operating over a large urban area may lose connectivity intermittently due to interference caused by building shadowing which makes the central controller unreachable leaving the swarm out of coordination.

This article presents TRUAV, a distributed multi-agent reinforcement learning framework based on independent tabular Q‑learning agents that eliminates the need for centralized training while maintaining comparable network performance. The design rests on three pillars. First, each UAV independently learns adaptive trajectories through local tabular Q‑learning using only locally observable information: nearby vehicle density, packet queue states at connected vehicles, neighbor UAV positions, and distance to the nearest RSU. This eliminates global state aggregation and reduces inter-agent communication to lightweight beacons. Second, a potential-game-inspired reward structure implicitly coordinates spatial distribution and discourages UAV clustering without explicit coordination messages. Third, a routing-demand-aware reward term directly couples trajectory decisions to forwarding utility by rewarding successful packet relays and penalizing excessive energy consumption. 

The remainder of this article is organized as follows. Section~\ref{sec:related} reviews related work and positions our contributions. Section~\ref{sec:system} describes the system model. Section~\ref{sec:methodology} presents the distributed reinforcement learning framework and reward design. Section~\ref{sec:simulation} details simulation parameters. Section~\ref{sec:results} presents performance comparisons. Section~\ref{sec:insights} distills IoT-oriented design insights, and Section~\ref{sec:conclusion} concludes with future directions.

\section{Related Work and Motivation}
\label{sec:related}

Research on UAV-aided vehicular networks has progressed through three main phases: static placement optimization, dynamic trajectory planning, and learning-based adaptive control. Understanding the contributions and limitations of each phase is essential for positioning the proposed framework within the broader research and IoT landscape.

\textbf{Static placement optimization.}
Jeong et al.~\cite{2.1} investigated static UAV deployment with integrated user association and bit allocation for UAV‑mounted cloudlets in mobile edge computing. Cho et al.~\cite{2.2} further demonstrated that static UAV deployment fails to cope with spatio-temporal hotspot dynamics  proposing a predictive path planning scheme for multiple UAVs to track time-varying network traffic under deadline constraints. Despite their analytical elegance, these placement-centric designs cannot adapt to time-varying vehicular traffic: UAVs positioned for rush-hour congestion remain idle off-peak, while shifting high-density road segments experience coverage and routing gaps. 

\textbf{Dynamic trajectory planning.}
Li~et~al.~\cite{1.5} proposed COACH, a collaborative joint trajectory and caching scheme for UAV-assisted edge caching networks, in which multiple cache-enabled UAVs act as flying containers under a multiagent deep reinforcement learning framework built on MADDPG, trading off user QoE, content updating cost, and dynamic hotspot coverage without requiring a pre-known content demand model. Zhou et al.~\cite{2.3} advanced this by incorporating pre-planned UAV trajectory knowledge into routing decisions, modeling packet forwarding as a time-dependent graph optimization to jointly minimize end-to-end delay and power consumption in dynamic UAV swarm networks. However, both approaches rely on offline path design derived from historical traffic patterns and cannot respond effectively to unexpected events such as accidents or road closures.

	\begin{figure*}[!t]
	\centering
	\includegraphics[width=\textwidth]{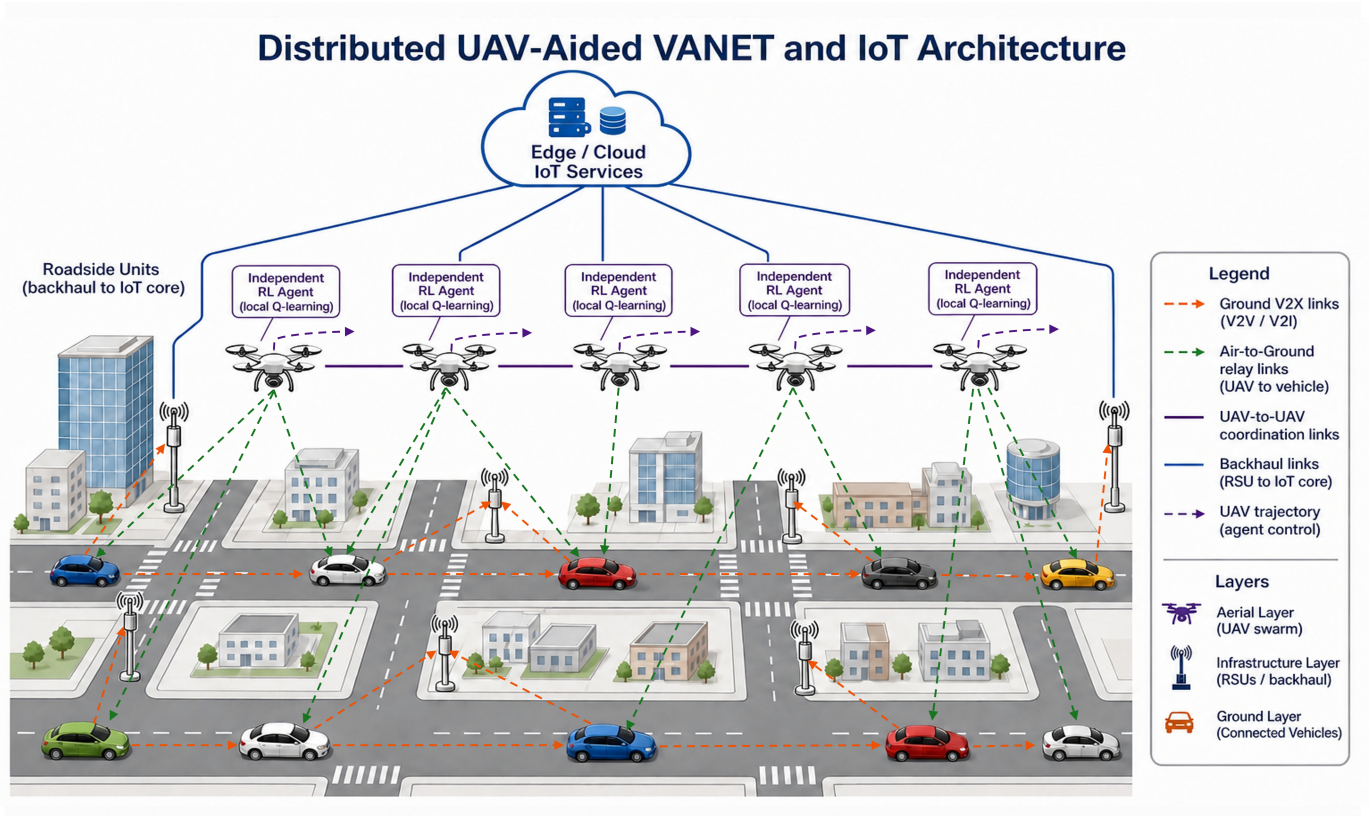}
	\caption{Distributed UAV-aided VANET and IoT architecture with independent reinforcement-learning agents per UAV.}
	\label{fig:architecture}
\end{figure*}

\textbf{Learning-based adaptive control.}
Shen et al.~\cite{1.6} proposed a deep reinforcement learning method with hierarchical prioritized experience replay for UAV collision avoidance in joint operational airspace. Ji~et~al.~\cite{1.7} proposed a  multi-agent deep reinforcement learning to jointly optimize uplink--downlink decoupled user association and rate-splitting multiple access beamforming in multi-UAV cellular networks. Zhan et al.~\cite{2.4} jointly optimized UAV deployment positions and dependent task offloading in a multi-UAV MEC network using a graph-attention-enhanced DRL framework, demonstrating how learning-based approaches can simultaneously handle placement and resource allocation for IoT-connected ground users. Chen et al.~\cite{2.5} proposed the closest related approach: a centralized multi-agent soft actor-critic framework that jointly optimizes trajectories and routing. Their global critic evaluates joint actions across all UAVs, demonstrating significant throughput improvement over static placement, but at the cost of high communication overhead and limited scalability.

\textbf{Distributed multi-agent learning.}
Hu et al.~\cite{2.6} proposed Graph Soft Actor-Critic for large-scale distributed multirobot coordination, leveraging graph neural networks to design distributed policies with strong zero-shot generalization in continuous action spaces. Tu et al.~\cite{2.7} introduced an adaptive role-learning framework integrating multiagent reinforcement learning with evolutionary algorithms, enabling UAVs and vehicles to dynamically refine task preferences without relying on predefined static structures. Lv et al.~\cite{2.8} demonstrated multi-agent reinforcement learning for UAV swarm communications against jamming, where shared observations and experience replay across neighboring UAVs stabilize distributed policy learning and improve sample efficiency.

Overall, static placement methods provide analytical tractability but lack adaptability. Trajectory-planning approaches introduce dynamics but remain constrained by pre-computation. Centralized deep reinforcement learning enables joint optimization but suffers from scalability limitations due to global state aggregation and control-plane load. Distributed multi-agent learning offers scalability, but has not been previously applied to the specific challenge of routing-aware UAV trajectory adaptation in VANETs. TRUAV fills this gap by combining independent, tabular Q-learning with a reward structure that aligns local decisions with global routing utility. From an IoT perspective, these works collectively highlight the tension between global optimality and deployment scalability when coordinating large ensembles of mobile IoT nodes such as connected vehicles and UAV relays.
	
	\section{System Model and Problem Formulation}
	\label{sec:system}
	
	We consider a UAV‑aided IoT‑enabled VANET deployed over an urban region in a smart city. The system follows a heterogeneous air–ground architecture in which connected vehicles constitute the ground layer, roadside units (RSUs) form the roadside infrastructure layer, and unmanned aerial vehicles (UAVs) provide an agile aerial relaying layer. Vehicles navigate through lane-constrained roads while communicating with other nearby nodes through short-range V2X links, leading to a series of multi-hop flows for safety and traffic management. Selected intersections and major corridors are installed with RSUs which offer access to edge/cloud IoT services as well as wired backhaul connectivity. The UAVs fly at a fixed low altitude and act as a flying gateway or relay, forming a line‑of‑sight air–ground link to the ground nodes to fill coverage gaps, facilitate inter‑cluster data forwarding, and assisting routing when vehicle connectivity between clusters is poor. The architecture of the heterogeneous air–ground system is shown in Figure~\ref{fig:architecture}.
	
	\subsection{Vehicle Mobility Model}	
	Vehicle motion in urban traffic is modeled using a standard car-following model (CFM) under which each vehicle accelerates, decelerates, and changes lanes. Road networks are represented as a regular grid of two-lane roads. Vehicles at an intersection probabilistically select a new direction using turning probabilities typical of urban areas. This setup yields realistic clustering behaviour during busy periods, with higher vehicle densities near major intersections and fewer vehicles in residential areas.
	
	\subsection{UAV Communication Model}	
	Communication links from UAVs to vehicles are modelled as probabilistic LoS where the probability depends on the elevation angle. Furthermore, the path loss consists of the free-space component plus an additional term accounting for attenuation due to urban clutter. For every link, the applicable signal-to-noise ratio is calculated using factors such as transmit power, antenna gains, path loss, noise power, and bandwidth, and this is then used to obtain an achievable data rate using the Shannon capacity relationship. Coordination over-the-air UAV-to-UAV links occurs in the same spectrum band, however, the LoS probability is much higher because of the aerial vantage point.  It is assumed that each UAV remains within a communication range in which vehicles can form a reliable data link to relay packets.
	
	\subsection{Energy Consumption Model}
	The energy consumption of each UAV is dominated by aerodynamic power for flying and RF power for communicating. In flight, we distinguish hovering from horizontal cruising. Hovering requires a base power to counter gravity. Horizontal flight needs additional propulsion power above this base level, and that extra power depends on frame characteristics and speed. To communicate, each UAV spends energy to transmit and receive packets over air–ground and air–air links. Per-packet costs depend on the UAV’s transmit power, modulation and coding scheme, and packet duration. We model these effects via mode-dependent power profiles: each UAV has a nominal hovering power level, an incremental propulsion power when in motion, and a communication power term that scales with active transmission time. The cumulative energy consumption along a trajectory is obtained by integrating the corresponding power profile over time and adding the communication energy of all forwarded packets, highlighting the trade-off between more aggressive trajectories that enhance routing utility and the limited on-board battery capacity.
	
	\subsection{Problem Statement}
	Our objective is to determine UAV trajectories that maximize network-wide routing utility under the aforementioned communication and energy constraints. The routing utility jointly captures four metrics: coverage (the percentage of vehicles within communication range of at least one UAV), packet delivery ratio (the fraction of generated packets successfully forwarded to an RSU), average relay delay, and total throughput. The key challenge is to achieve this optimization in a distributed manner, without centralized coordination or global state knowledge, and without overloading the IoT control plane with inter-UAV coordination traffic.
	
	\section{Proposed TRUAV Framework}
	\label{sec:methodology}
	
\begin{figure}[!ht]
		\centering
		\includegraphics[width=\columnwidth]{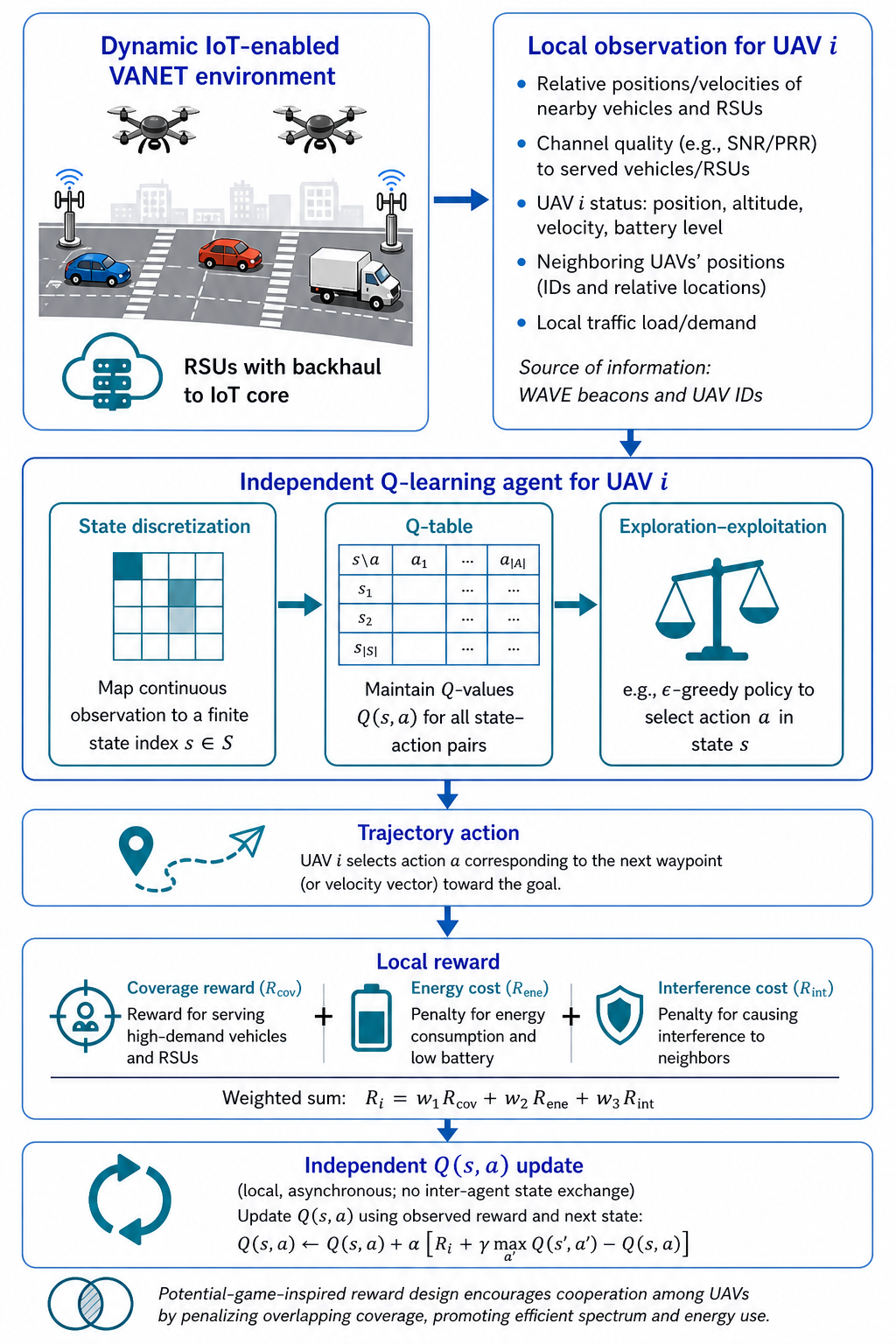}
		\caption{Logical workflow of the proposed TRUAV framework for UAV \textit{i}.}
		\label{fig:logicalworkflow}
	\end{figure}

	The TRUAV framework departs from centralized deep reinforcement learning approaches by assigning an independent tabular Q‑learning agent to each UAV. Each agent observes only its local environment, selects movement actions, receives rewards based on relay performance and coverage, and updates its Q‑table without exchanging state information with other agents or a central controller. The logical workflow of the proposed framework is illustrated in Figure~\ref{fig:logicalworkflow}.	
	
	\subsection{Local State Representation}
	
	The state vector for each UAV captures four locally observable categories:
	\begin{itemize}[leftmargin=*,nosep]
		\item Vehicle density within communication range, indicating forwarding demand.
		\item Aggregate packet queue length at connected vehicles, measuring backlog urgency.
		\item Relative positions of neighboring UAVs, enabling spatial separation to prevent clustering.
		\item Distance to the nearest RSU, informing end-to-end delivery feasibility.
	\end{itemize}
	
	The agent requires no communication with others to obtain these observations. Periodic WAVE beacons help estimate vehicle density and queue lengths. RSU locations are known prior, and the positions of neighboring UAVs are obtained from the aerial identification broadcasts. The state is encoded as a compact tuple formed by a fine spatial grid together with ordinal levels for density, queue load, neighbor distance. This discretization yields a finite state space compatible with tabular Q-learning, allowing expedient onboard updates without the need for neural network inference.
	
	\subsection{Action Space and Selection}	
	Each UAV can select from 16 different actions. Specifically, it can choose from eight possible flight directions at 45-degree intervals and two speed levels (half or full maximum). In this design, agility is balanced against learning complexity. Since all UAVs fly at a constant altitude, vertical motion is not considered, which facilitates regulation and simplifies the learning problem. The agent employs an exploration-exploitation strategy to choose an actions: at the start of training it selects random action with high probability, and as the training progresses, it starts to select the action with a higher Q-value.
	
	\subsection{Reward Function Design}	
	We design the local reward so that each UAV maximizes a quantity aligned with network-wide routing utility. For UAV \(i\) at time step \(t\), the reward is defined as
	\begin{equation}
		\label{eq:reward}
		r_i(t)=\alpha R_i^{\mathrm{relay}}(t)+\beta R_i^{\mathrm{cov}}(t)-\gamma R_i^{\mathrm{eng}}(t)-\delta R_i^{\mathrm{ov}}(t),
	\end{equation}
	where \(R_i^{\mathrm{relay}}(t)\), \(R_i^{\mathrm{cov}}(t)\), \(R_i^{\mathrm{eng}}(t)\), and \(R_i^{\mathrm{ov}}(t)\) denote the relay, coverage, energy, and overlap terms, respectively. The weights are set to \(\alpha=1.0\), \(\beta=0.25\), \(\gamma=0.15\), and \(\delta=0.20\), so that delivering packets is favored, useful coverage is rewarded, and high energy use or redundant overlap with other UAVs is discouraged. These weight values were chosen based on manual tuning and a brief sensitivity sweep to balance packet delivery, coverage, and energy consumption.
	
	\subsection{Distributed Q-Learning with Asynchronous Updates}
	
	Each UAV maintains an independent tabular Q‑table  initialized with small random values. During each time step, every UAV executes an identical learning cycle: observe local state, select an action according to the current exploration–exploitation strategy, execute movement, compute reward based on relays, coverage, and energy, observe the next state, and update the corresponding Q-value entry. The learning rate controls the sensitivity to new experience, while a discount factor balances immediate rewards against long-term positioning value. Asynchronous, independent updates naturally accommodate stochastic vehicle and packet dynamics without requiring synchronized coordination.
	
	\subsection{Potential-Game-Inspired Coordination}
	
	We interpret the multi-UAV interaction through the lens of potential games. In such games, any agent’s utility change from a unilateral action can be related to the change in a global potential function. In TRUAV, the reward structure in (\ref{eq:reward}) is designed so that the global objective of maximizing routing utility is closely aligned with the sum of individual rewards, while the overlap term discourages excessive spatial overlap between neighboring UAVs. This alignment allows the tabular Q‑learning updates at each agent to steer the swarm toward configurations that are close to optimal for the overall routing utility. When a UAV enters a region already covered by another, this regularization reduces its reward, creating an implicit repulsive force that spreads the swarm without explicit coordination messages. Each UAV experiences lower rewards in densely covered regions and gradually learns to avoid them.
	
	Prior work shows that decentralized Q-learning converges to a stable configurations under standard assumptions on the learning rates and the potential-game structure~\cite{2.7}.

		\begin{figure*}[!t]
		\centering
		\subfloat[Network coverage]{\includegraphics[width=.35\linewidth, height=4cm]{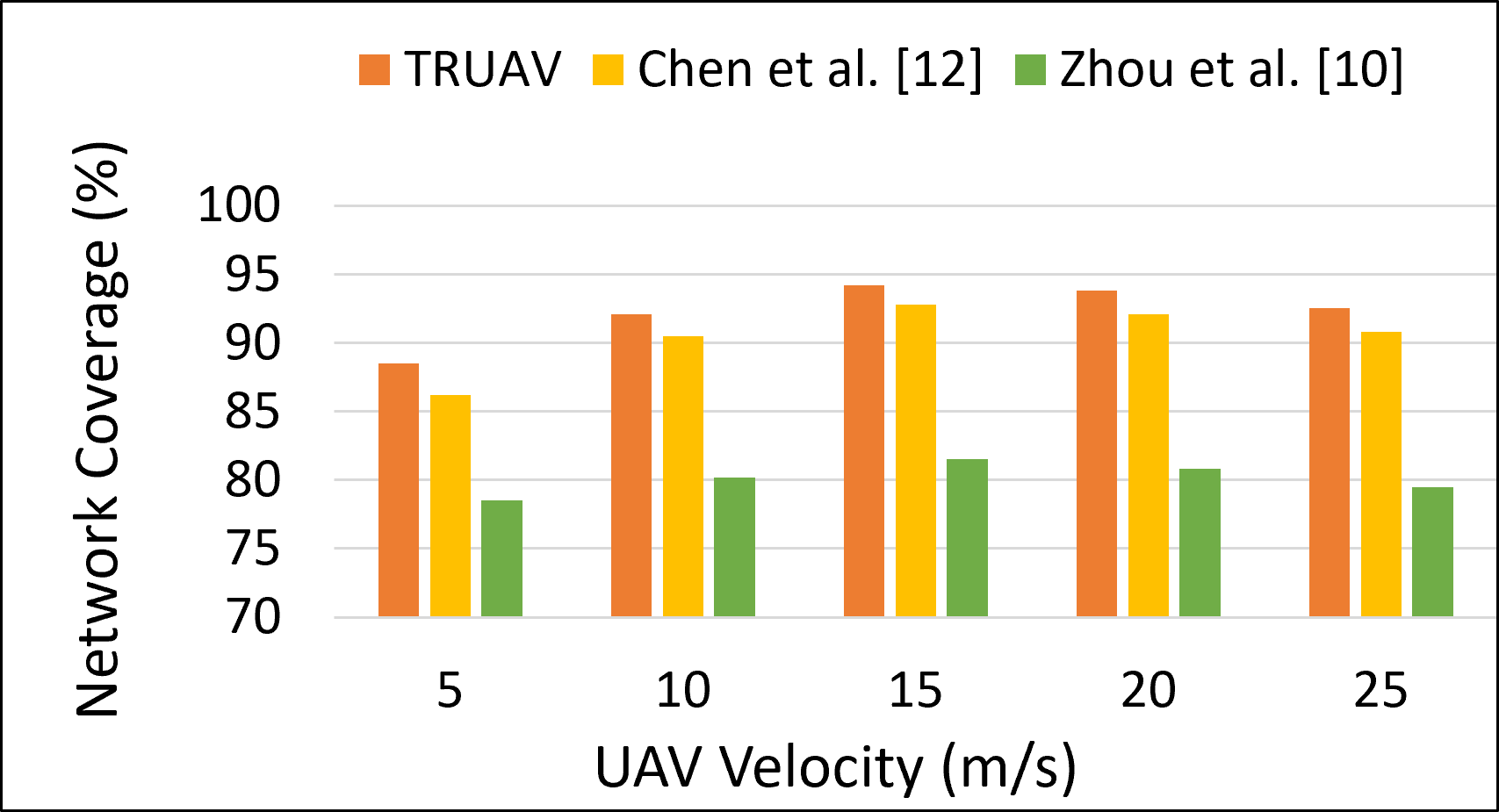}}\quad
		\subfloat[Average relay delay]{\includegraphics[width=.35\linewidth, height=4cm]{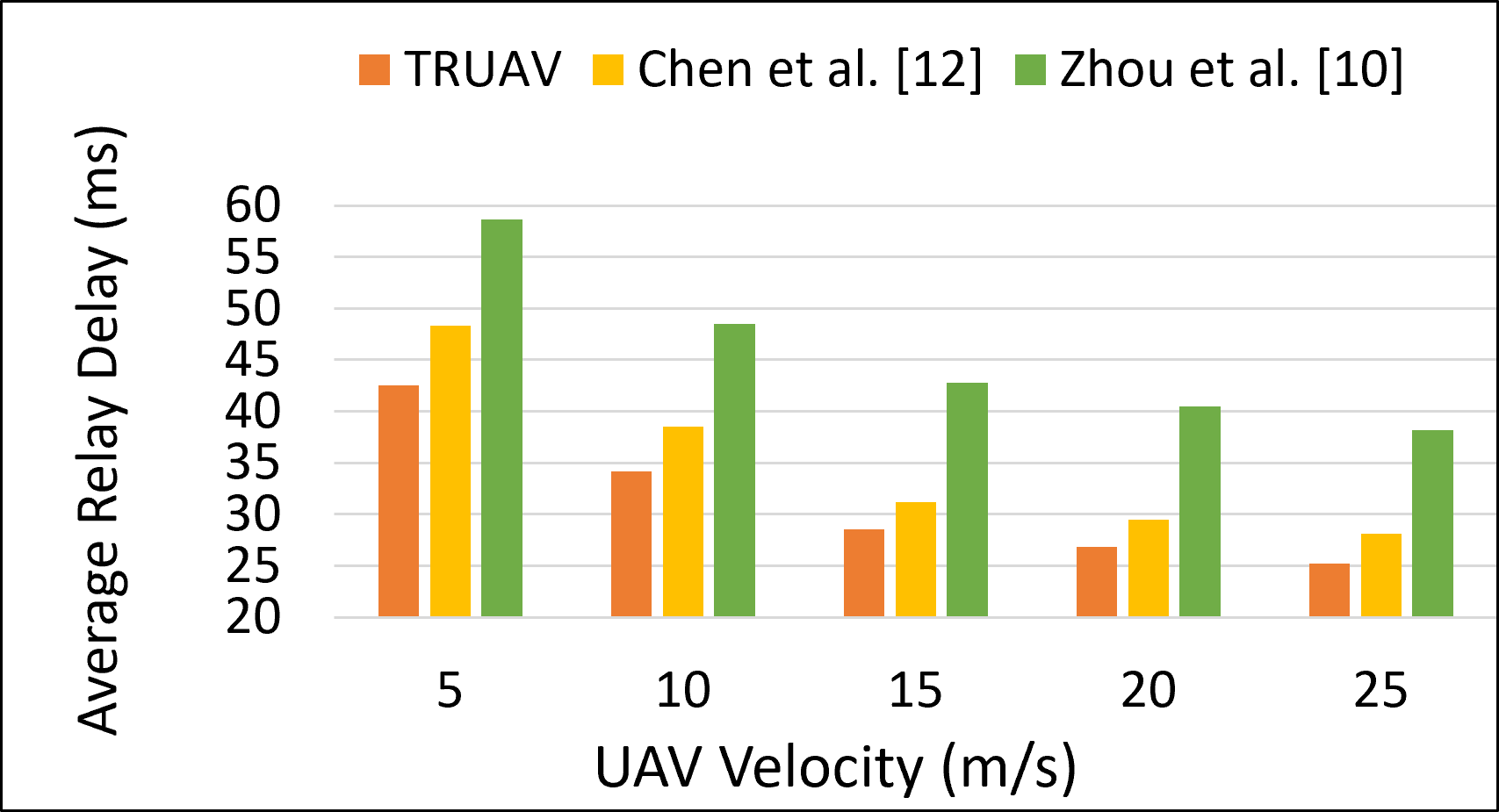}} \\
		\subfloat[Energy consumption]{\includegraphics[width=.35\linewidth, height=4cm]{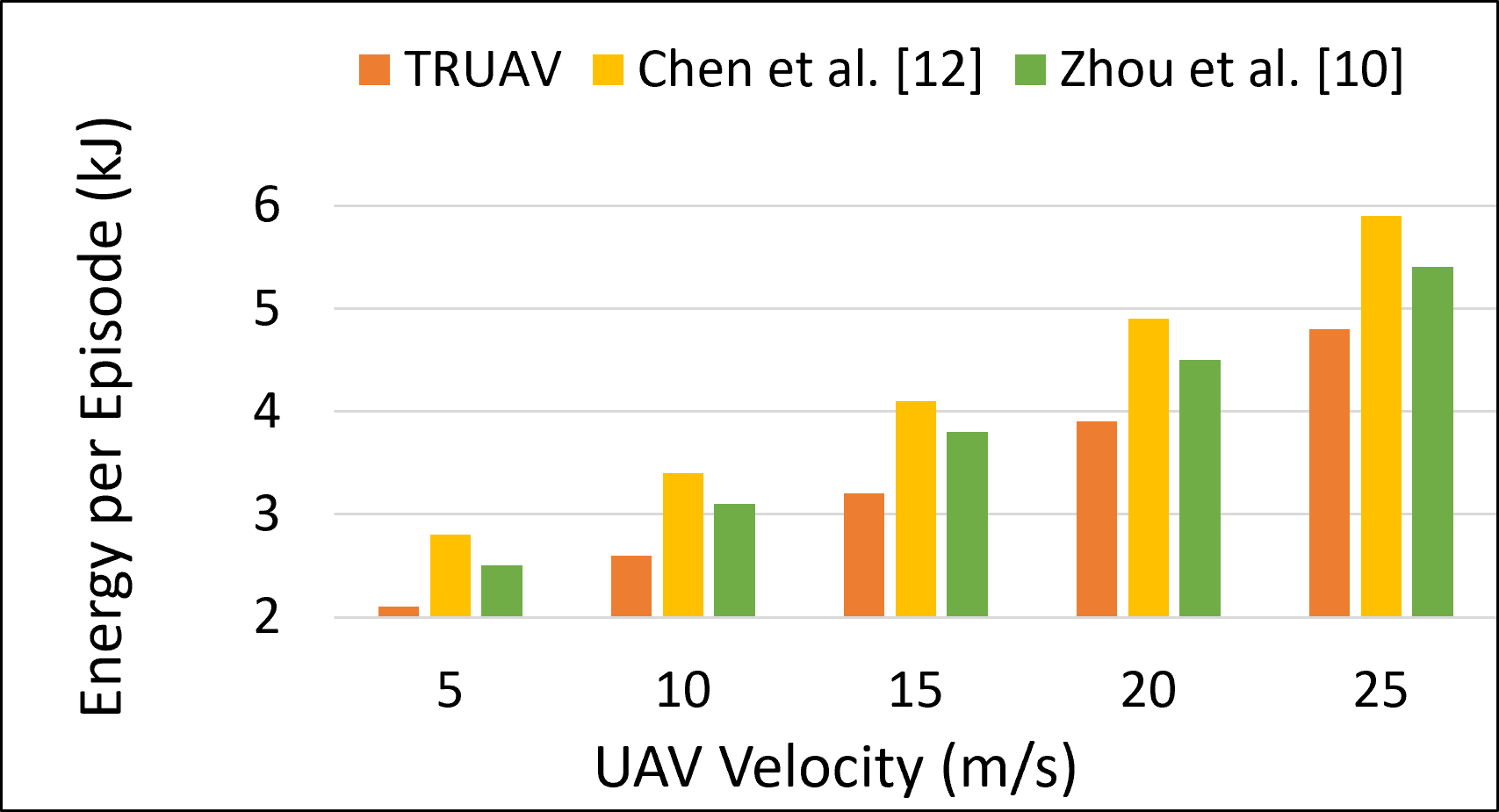}}\quad
		\subfloat[Packet delivery ratio]{\includegraphics[width=.35\linewidth, height=4cm]{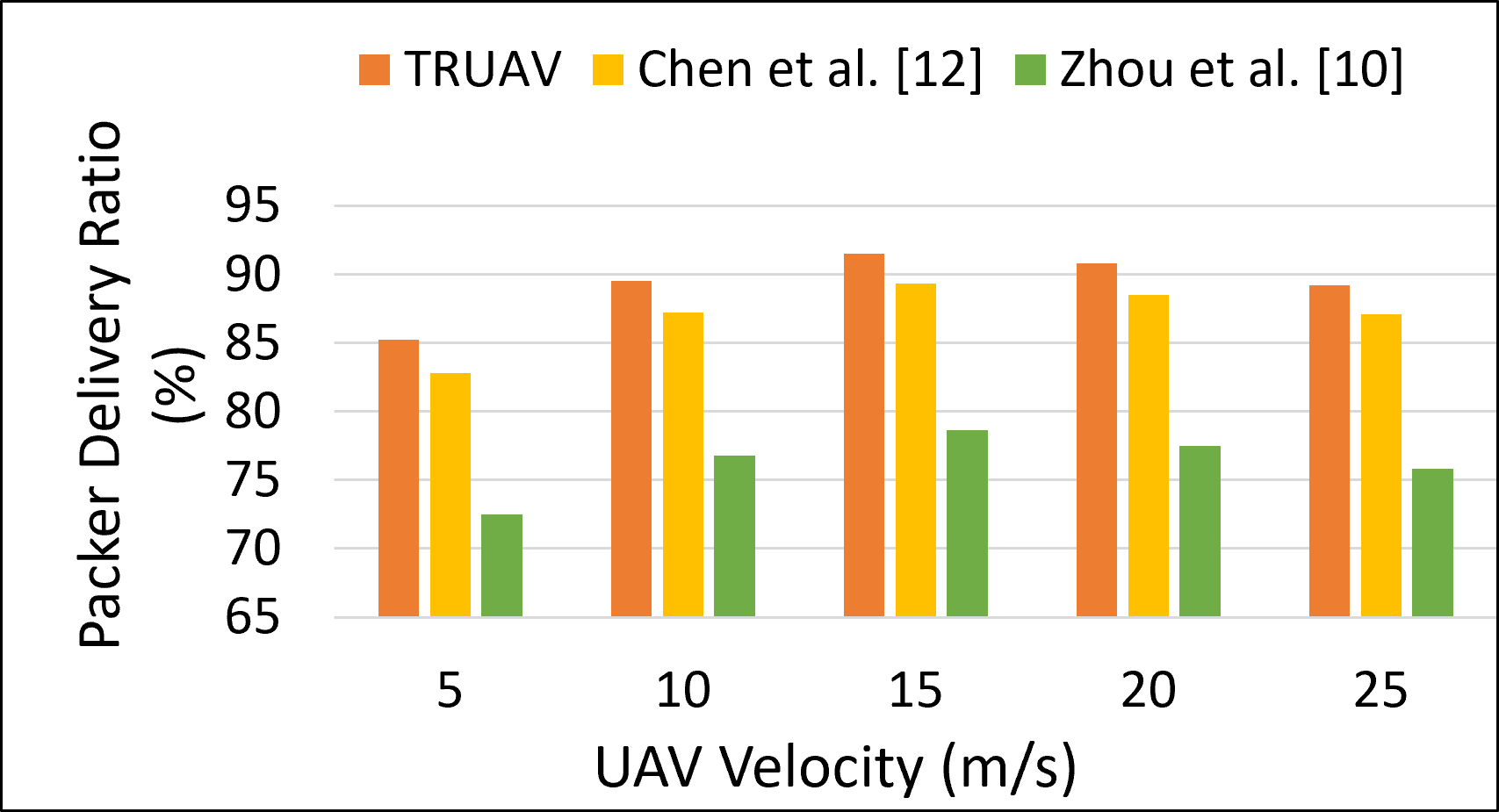}}
		\caption{Performance versus UAV velocity}
		\label{fig:PUV}
	\end{figure*}
	
	\section{Simulation Setup and Parameters}
	\label{sec:simulation}
	
	We implement TRUAV and both baseline schemes in Python using NumPy and Matplotlib. The simulation environment instantiates the system model described in Section~\ref{sec:system} over a $5{,}000$\,m by $5{,}000$\,m urban area populated by ground vehicles, UAV relays, and RSUs. Vehicles follow the car-following mobility model on a grid road network, while UAVs operate at a fixed altitude and move according to the selected trajectory control policy. RSUs are placed near the corners of the region to provide backhaul connectivity.
	
	Communication follows IEEE 802.11p in the 5.9\,GHz band with a 20\,MHz channel, and the probabilistic line-of-sight and path-loss assumptions of Section~\ref{sec:system} are used to determine link rates. UAV energy parameters reflect commercial quadcopter profiles and are summarized in Table~\ref{tab:parameters}. It also lists the main environment, communication, and reinforcement learning hyperparameters used in all experiments.
	
	\begin{table}[!t]
		\centering
		\caption{Simulation Parameters}
		\label{tab:parameters}
		\footnotesize
		\begin{tabular}{@{}l@{}c@{\hspace{0.5em}}p{4cm}}
			\toprule
			\textbf{Parameter} & \textbf{Value} & \textbf{Description} \\
			\midrule
			\multicolumn{3}{@{}l}{\textit{Environment}} \\
			Simulation area    & 5000 by 5000 m & Urban coverage region \\
			Number of vehicles & 200            & Ground vehicles (CFM mobility) \\
			Number of UAVs     & 5              & Aerial relay platforms \\
			Number of RSUs     & 4              & Corner-mounted backhaul nodes \\
			UAV altitude       & 100 m          & Fixed flight level \\
			\midrule
			\multicolumn{3}{@{}l}{\textit{UAV Movement}} \\
			Speed range        & 5--25 m/s      & UAV horizontal velocity \\
			Communication range& 500 m          & UAV-to-vehicle link range \\
			Hover power        & 200 W          & Stationary power draw \\
			Move power         & 250 W          & Propulsion power draw \\
			\midrule
			\multicolumn{3}{@{}l}{\textit{Communication}} \\
			Bandwidth          & 20 MHz         & IEEE 802.11p channel \\
			Noise power        & $10^{-9}$ W    & Thermal noise floor \\
			Vehicle transmit power & 0.1 W      & Vehicle-to-UAV transmit power \\
			UAV transmit power & 1.0 W          & UAV-to-UAV and UAV-to-RSU transmit power \\
			\midrule
			\multicolumn{3}{@{}l}{\textit{Reinforcement Learning}} \\
			Learning rate      & 0.1            & Q-value update step size \\
			Discount factor    & 0.95           & Future reward weight \\
			Initial exploration rate & 1.0      & Exploration rate at start \\
			Final exploration rate   & 0.05     & Exploration rate at end \\
			Episodes           & 5000           & Training iterations \\
			Steps per episode  & 100            & Simulation steps \\
			\bottomrule
		\end{tabular}
	\end{table}
	
	The baseline approaches implemented for comparison are the centralized multi-agent soft actor–critic framework proposed by Chen et al.~\cite{2.5} and the trajectory-knowledge routing protocol developed by Zhou et al.~\cite{2.3}. The centralized baseline maintains a global critic that takes the concatenated state of all UAVs as input and outputs joint actions for the entire swarm. The trajectory-knowledge baseline follows pre-planned patrol routes centered at major road intersections, with routing decisions made using advance knowledge of UAV positions along these routes.
	
	Each simulation run spans $5{,}000$ training episodes, with each episode consisting of $100$ time steps representing $100$ seconds of simulated operation. Metrics are recorded at the episode level and averaged over sliding windows of $500$ episodes to produce smoothed convergence curves. All reported results correspond to the mean over $10$ independent simulation runs with different random seeds.

	\section{Performance Evaluation}
	\label{sec:results}
	
	\subsection{Impact of UAV Velocity on Network Coverage}
	
	Figure~\ref{fig:PUV}a shows coverage improving with UAV speed from 5 to 15 meters per second as faster repositioning allows UAVs to follow moving vehicle clusters more effectively. Beyond 15 meters per second, coverage declines slightly due to reduced dwell time over any given region. The trajectory-knowledge baseline provides the lowest coverage since fixed patrol paths cannot adapt to spatial traffic variations. At 15 meters per second, the proposed approach achieves roughly 94 percent coverage, exceeding the centralized deep reinforcement learning baseline by more than one percentage point.
	
	\subsection{Impact of UAV Velocity on Relay Delay}
	
	Figure~\ref{fig:PUV}b shows average relay delay decreasing monotonically with speed as faster UAVs respond more rapidly to changes in demand. The proposed approach achieves the lowest delay at every velocity: at 25 meters per second, TRUAV attains about 25.2 milliseconds average delay, compared with about 28.1 milliseconds for the centralized baseline and near to 40 milliseconds for the trajectory-knowledge scheme. The improvement is attributed to immediate local action selection without controller latency or global state aggregation.

	\subsection{Energy and Packet Delivery Ratio vs. Velocity}
	
	Figure~\ref{fig:PUV}c shows episode-level energy consumption rising approximately linearly with UAV speed. The proposed framework consumes the least energy across the entire velocity range by eliminating global state broadcasts. At 25 meters per second, TRUAV consumes about 4.8 kilojoules per episode, representing approximately 19 percent less energy than the centralized baseline. As per Figure~\ref{fig:PUV}d, the packet delivery ratio achieves its maximum at $15$ m/s for all the schemes wherein TRUAV maintains the highest ratio for all the speeds with a peak around 91.5 percent.
	
	The proposed TRUAV method has an advantage over the baselines considered across key performance metrics at optimum velocity of $15$ m/s. The adaptive repositioning capability enables UAVs to track moving IoT traffic clusters, resulting in coverage and delivery improvements of several percentage points over the centralized and trajectory-knowledge baselines. The distributed framework is more energy efficient because it omits broadcasting the global state, resulting in notable gains relative to the centralized framework.
	
	\section{IoT-Oriented Design Insights}
	\label{sec:insights}
	
	TRUAV identifies three practical design principles for aerial IoT gateways and other connected vehicles from an IoT deployment perspective.
	
	\textit{Couple trajectories to traffic demand:}
	Rewarding successful packet relays rather than pure coverage ensures that learned flight paths serve actual communication demand. This is crucial in IoT scenarios where traffic patterns, not just geometry, determine the value of connectivity and the timeliness of information.
	
	\textit{Exploit implicit coordination:}
	Incorporating a coverage-regularization term into the reward function encourages UAVs to remain spatially separated and thus avoid cluttering. This is done without any explicit cooperation messages or sharing global state. This implicit coordination suits constrained IoT control channels and avoids single points of failure.
	
	\textit{Keep learning lightweight:}
	Using a compact, discretized states representation makes distributed reinforcement learning feasible without deep neural networks. This approach is attractive for resource-constrained UAV platforms and limited IoT backhaul links and improves robustness to channel conditions and physical hardware.

\section{Conclusion and Future Work}
\label{sec:conclusion}
This article presented TRUAV, a multi-agent reinforcement learning-based distributed framework for joint UAV trajectory planning and routing enhancement in UAV aided, IoT-enabled VANETs. The framework’s decentralized Q-learning agents, which rely only on local observations of vehicle density, packet queue states, and neighbor positions, thereby eliminating the scalability bottleneck of centralized training by avoiding global state aggregation and centralized control components. The proposed reward design, grounded in potential games with coverage regularization, has a principled effect of guiding selfish agents toward network-level beneficial configurations without requiring explicit coordination or state aggregation among agents. Simulation results show that the proposed protocol outperforms the evaluated baselines.

This work opens up several interesting avenues of research. Introducing federated-learning ideas, whereby UAVs periodically share distilled policy information instead of raw state information, could accelerate convergence while preserving operational privacy through reduced information entanglement and lower communication overhead. The framework can be extended to allow UAVs to optimize the LoS conditions created by varying building heights within urban canyons, potentially surpassing the coverage achievable from a fixed altitude. By combining the learning framework with predictive traffic models, UAVs could be positioned in anticipation of changing vehicle flows rather than only adapting after the fact, enabling proactive trajectory planning. 
Another direction is to enrich the state and reward design with additional QoS indicators, such as latency bounds for different traffic classes, to facilitate the integration of diverse IoT applications with varying reliability requirements. Finally, extending the framework to coordinate heterogeneous UAV fleets with different communication ranges, flight speeds, and energy capacities would better reflect realistic deployments involving multiple UAV types within an IoT system.

	\bibliographystyle{IEEEtran}
	\bibliography{sample-base}

\end{document}